\documentclass[11pt]{article}
\pdfoutput=1
\usepackage{jcappub}
\usepackage[T1]{fontenc} 

\usepackage{graphicx,amsfonts,amssymb,amsbsy}
\usepackage{amsmath,amsthm,latexsym}

\begin{document}

\title{Gravitational wave asteroseismology limits \\ from low density nuclear matter and perturbative QCD}

\author[a]{C. V\'asquez Flores}
\author[]{and}
\author[b]{G. Lugones}

\affiliation[a]{Universidade Federal do  Maranh\~ao, Departamento de F\'isica, Campus Universit\'ario do Bacanga, CEP 65080-805, S\~ao Lu\'is, Maranh\~ao, Brazil.}
\affiliation[b]{Universidade Federal do ABC, Centro de Ci\^encias Naturais e Humanas, Avenida dos Estados 5001- Bang\'u, CEP 09210-580, Santo Andr\'e, SP, Brazil.}

\emailAdd{cesarovfsky@gmail.com}
\emailAdd{german.lugones@ufabc.edu.br}

\abstract{
We investigate the fundamental mode of non-radial oscillations of non-rotating compact stars in general relativity using a set of equations of state (EOS) connecting state-of-the-art calculations at low and high densities. Specifically,  a low density model based on the chiral effective field theory (EFT) and high density results based on perturbative Quantum Chromodynamics (QCD) are matched through different interpolating polytropes fulfilling thermodynamic stability and subluminality of the speed of sound, together with the additional requirement that the equations of state support a two solar mass star. We employ three representative models (EOS I, II and III) presented in Ref. \cite{2014ApJ...789..127K} such that EOS I gives the minimum stellar radius, EOS II the maximum stellar mass, and EOS III the maximum stellar radius.
Using this family of equations of state, we find that  the frequency and the damping time of the $f$-mode are constrained within narrow quite model-independent windows. We also analyze some proposed empirical relations that describe the $f$-mode properties in terms of the average density and the compactness of the neutron star. We discuss the stringency of these constrains and the possible role of physical effects that cannot be encoded in a mere interpolation between low and high density EOSs. 
}


\keywords{neutron stars, gravitational waves / sources}
\maketitle
\flushbottom

\section{Introduction}

The recent detection of gravitational waves (GWs) from a binary neutron star (NS) merger by the LIGO-Virgo collaboration \cite{abbott2017b} opens a new window for exploring the equation of state (EOS) of dense matter at NS interiors. 
Several properties directly related to the EOS can be inferred from the gravitational wave signal of the merging event, such as the masses of the two coalescing objects \cite{abbott2017b}. 
Further details on the EOS can be obtained from the tidal deformability during late inspiral \cite{flanagan2008,read2009} and from the observation the  electromagnetic counterpart  associated with short gamma-ray bursts \cite{abbott2017c} or with the transient optical-near-infrared source powered by the synthesis of large amounts of very heavy elements via rapid neutron capture \cite{pian2017}. 
Additionally, lot of information can be obtained from neutron star oscillations both before merging and in the post-merger phase. 
In fact, neutron stars can be tidally deformed by the orbital motion which may in principle excite NS oscillations; but, in general NS normal modes have much higher frequency than the orbital frequency of inspiral binary systems. Nonetheless, in generic elliptic orbits, resonant oscillations at a frequency much higher than the frequency scale set by the inverse of the orbital period are possible, in particular if the NS companion is a black hole \citep{shibata1994,parisi2017}. 
In the post-merger phase, the violent dynamics of the merging lefts (if a direct collapse into a black hole doesn't occur) a massive NS which is strongly oscillating  (see e.g. \cite{bauswein2016,andersson2011,faber2012,rezzolla2013}). In particular, the fundamental quadrupolar fluid mode of the remnant is strongly excited and dominates the post-merger GW signal. 
Pulsation modes can also be  excited in newly born compact objects associated with the violent dynamics of core collapse supernovae \cite{Camelio:2017nka}, due to starquakes and glitches \cite{Warszawski:2012zq},  by accretion in a binary system, or by the rearrangement of the star following the conversion of a hadronic star into a quark star \citep{lugones2002,abdikamalov2008}.

Neutron star oscillations can be analyzed within the quasi-normal mode formalism. An adequate base for a rigorous treatment of this formalism  in general relativity was depicted in the pionering work of Thorne and Campolattaro \cite{1967ApJ...149..591T,1968ApJ...152..673T}. The theory  was later completed by Lindblom and Detweiler, who brought the analytic framework to a form suitable for the numerical integration of the equations \cite{1983ApJS...53...73L}. 
Amongst all the possible modes of oscillation,  the fundamental ($f$) mode is very important, because it can be excited easier than other high frequency modes, and as a consequence it could be observed by the present GW detectors.

At present, a trustworthy determination of the EOS of strongly interacting matter at densities above few times the nuclear saturation density  is still a challenge.  The EOS can be reliably obtained up to the nuclear saturation density, but for larger densities, its determination depends crucially on the knowledge of strong interactions in a regime that cannot be reached experimentally.  As a consequence, there is a large number of phenomenological high-density EOSs  in the literature that incorporate several aspects that could play a crucial role at the inner core of the star, such as three-body forces, bosonic condensates, hyperonic degrees of freedom and quark  matter. 
On the other hand, efforts have been made to  constrain the EOS and the structure of cold neutron stars from first principles (see e.g. \cite{2014ApJ...789..127K,2013ApJ...773...11H}). This can be achieved by connecting by some means an EOS describing the well understood regime of low densities described by a low-energy chiral effective field theory \cite{2013ApJ...773...11H} with the ultrahigh density regime described by perturbative  quantum chromodynamics (QCD) \cite{2014ApJ...781L..25F}.  
In this work we use this kind of EOSs to study the resulting effects on the \textit{f}-mode of NSs. 

The paper is organized as follows: in section \ref{EOS} we introduce the EOSs. In section \ref{formalism} we summarize the formalism for studying non-radial oscillations of non-rotating compact stars in general relativity. In section \ref{results} we present our results and  in section \ref{conclusions} our main conclusions.

\section{Equations of state} \label{EOS}

In the outer layers of a compact object, the matter composition goes from nuclei immersed in an electron sea in the outer stellar crust to increasingly neutron-rich matter in the inner crust and outer core. The EOS in this regime has been usually obtained through many-body calculations employing phenomenological potentials, typically accounting for two- and three-nucleon interactions (see e.g.~\cite{1998PhRvC..58.1804A}). 
More recently,  the development of the chiral effective field theory  (EFT) has given a frame for a systematic expansion of nuclear forces at low momenta explaining the hierarchy of two-, three-, and weaker higher-body forces \citep{2013PhRvC..87a4322C,2012PhRvC..85c2801G,2013PhRvC..87a4338H,2010PhRvC..82a4314H,2012PhRvC..86e4317S,2013PhRvL.110c2504T}. At present, microscopic calculations based on chiral EFT interactions allow a reliable determination of  the properties of neutron star matter up to the nuclear saturation density $n_0$;  specifically,  the pressure is currently known to roughly $\pm 20\%$ accuracy at $n_0$ \citep{2013PhRvL.110c2504T,2010PhRvC..82a4314H}. In the regime of intermediate densities above $n_0$, our understanding of the EOS is insufficient, but strong constraints can be imposed from precisely determined neutron star masses  \cite{demorest2010,antoniadis2013}.

At high densities, the EOS of cold quark matter can be obtained through perturbative QCD. Calculations have first determined the EOS for massless quarks to order $\alpha_s^2$ in the strong coupling constant  \cite{1977PhRvD..16.1169F,1978PhRvD..17.2092B},  and later generalized to systematically include the effects of a nonzero strange quark mass at two \citep{2005PhRvD..71j5014F} and three loops \citep{2010PhRvD..81j5021K}.
The EOS depends on an unphysical parameter, the scale of the chosen renormalization scheme. This dependence, which diminishes order by order in perturbation theory, offers a convenient way to estimate the contribution of the remaining, undetermined orders, and thus serves as a quantitative measure of the inherent uncertainty in the result \cite{2014ApJ...789..127K}.  Here we use the numerical EOS derived in \cite{2010PhRvD..81j5021K} which is well represented by a fitting function for the pressure in terms of the baryon chemical potential $\mu_B$   \cite{2014ApJ...781L..25F}. Fixing the strong coupling constant and the strange quark mass at arbitrary reference scales (using lattice and experimental data), the EOS of quark matter in $\beta$-equilibrium reads 
\begin{equation}
\label{EOS1}                                                                                                                                   
P_{QCD}(\mu_{B})=P_{SB} \left[    c_{1} - \dfrac{a(X)}{(\mu_{B}/GeV)-b(X)}   \right],
\end{equation}
with $P_{SB} = \tfrac{3}{4\pi^2}(\mu_{B}/3)^{4}$, $a(X) = d_{1}X^{-\nu_{1}}$ and $b(X) = d_{2}X^{-\nu_{2}}$, 
where $X$ is a parameter proportional to the renormalization scale of the theory (typically varied from $1$ to $4$). The parameters are  $c_{1} = 0.9008$,   $d_{1}= 0.5034$,   $d_{2}= 1.452$,    $\nu_{1}= 0.3553$ and $\nu_{2}= 0.9101$.
These formulae reproduce correctly the full three-loop pressure, the quark number density and the speed of sound to per cent accuracy for baryon chemical potentials up to $6$ GeV \cite{2014ApJ...789..127K,2014ApJ...781L..25F}.

To obtain the EOS for any density,  the following procedure is used. At densities below $1.1n_0$,  the chiral EFT EOS of \cite{2013PhRvL.110c2504T} is used, assuming that the true result is within the error band given in this reference. 
At baryon chemical potentials above 2.6 GeV, where the relative uncertainty of the quark matter EOS is as large as the nuclear matter one at $n=1.1 n_0$, the expression of Eq. \eqref{EOS1} is used. Between these two regions, it is assumed that the EOS is well approximated by an interpolating polytrope built from two monotropes of the form 
\begin{equation}
P_{i}(\mu_{B}) = k_{i}\left( n_{i}^{\gamma - 1} + \dfrac{\gamma_{i}-1}{k_{i}\gamma_{i}}(\mu_{B} - \mu_{B,i}) \right)^{\tfrac{\gamma_{i}}{\gamma_{i}-1}}, 
\end{equation}
where $i = 1,2$, with $\mu_{B,i}$ and $n_{i}$ being the baryon chemical potential and the baryon density respectively. 
These functions are connected smoothly and also allowing a density jump at the matching point of the two monotropes due to a first-order phase transition.

\begin{figure}[tb]
\centering
\includegraphics[angle=0,scale=0.35]{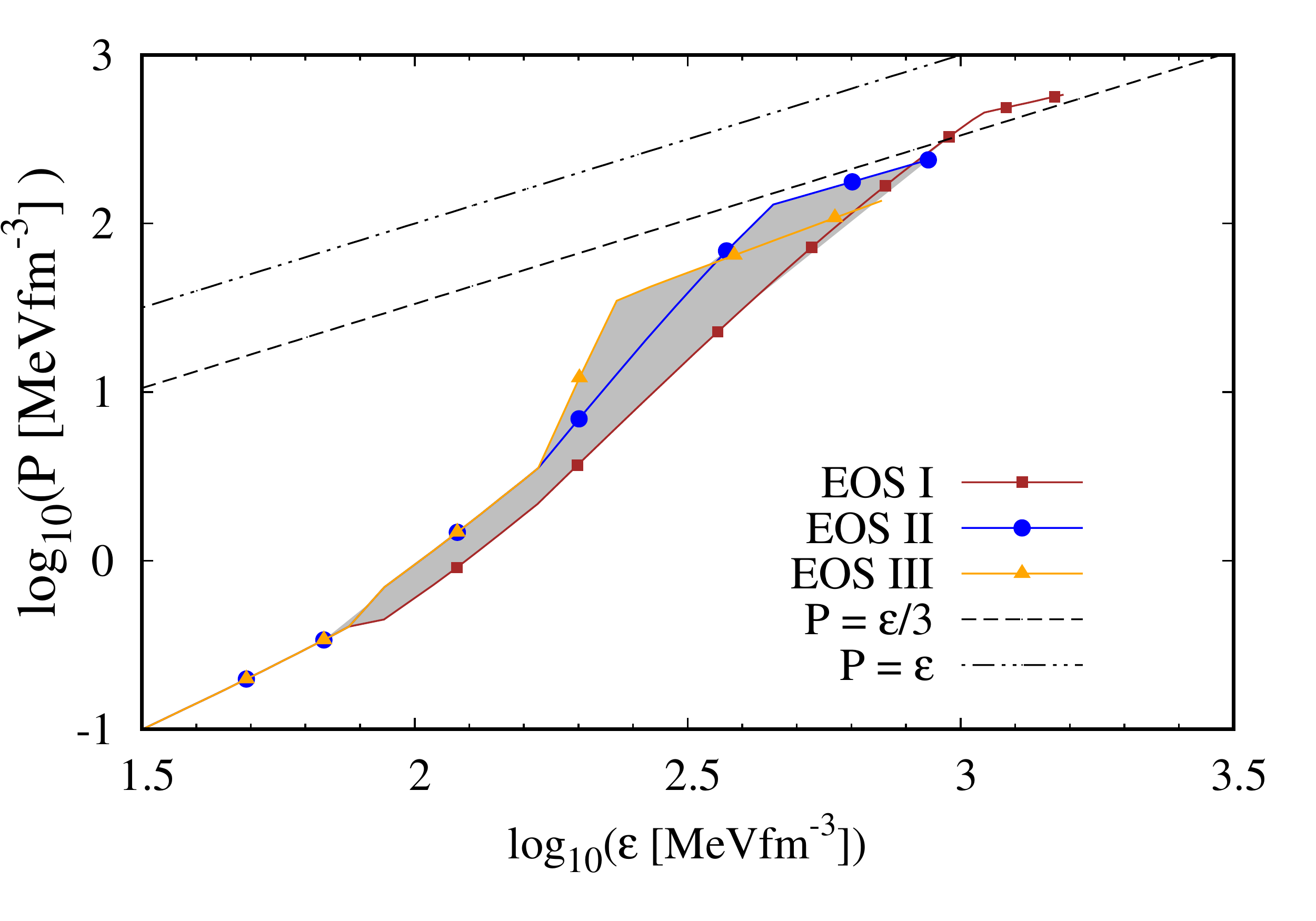}
\caption{Pressure $P$ versus energy density $\epsilon$ for three representative EOSs  that delimit a region of models that support a star of $2 M_{\odot}$.  These EOSs were chosen taking into account the mass-radius relationship they provide after the integration of the stellar structure  equations (see Sec. \ref{formalism}). In fact, EOS I gives the minimum radius, EOS II the maximum mass, and EOS III the maximum radius for compact stars.}
\label{fig1}
\end{figure}

\begin{figure*}[tb]
\centering
\includegraphics[angle=0,scale=0.24]{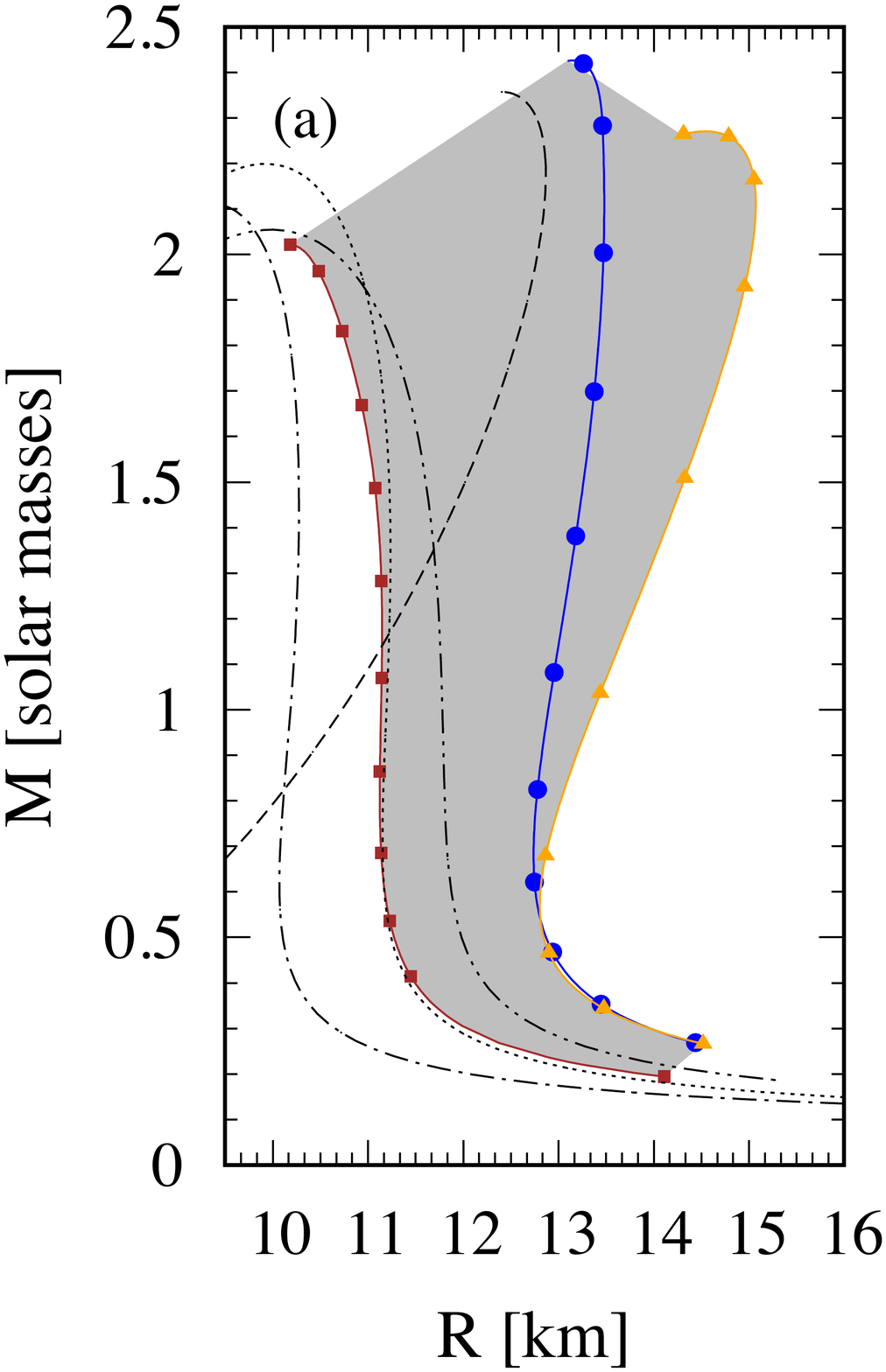} 
\includegraphics[angle=0,scale=0.24]{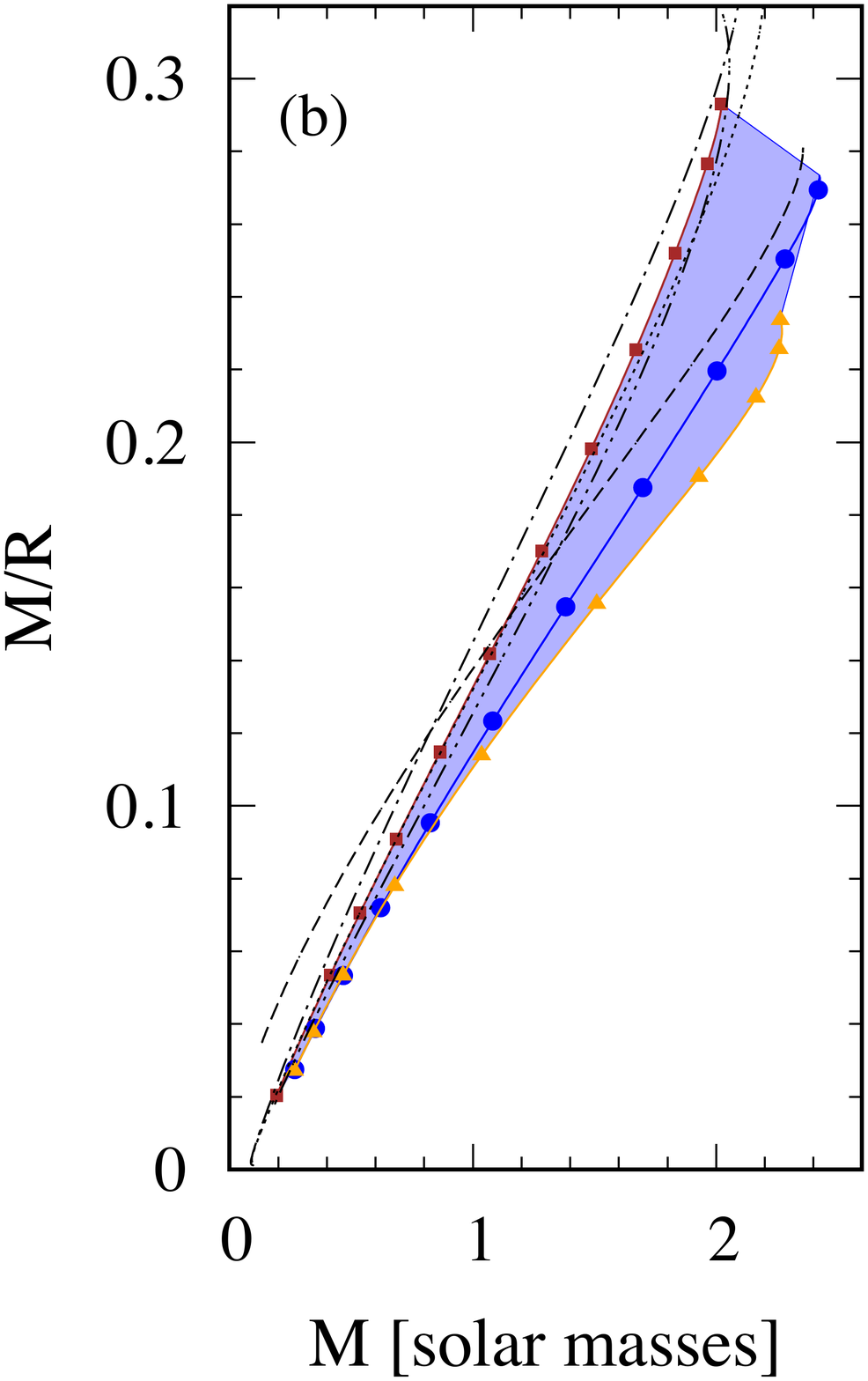} 
\includegraphics[angle=0,scale=0.24]{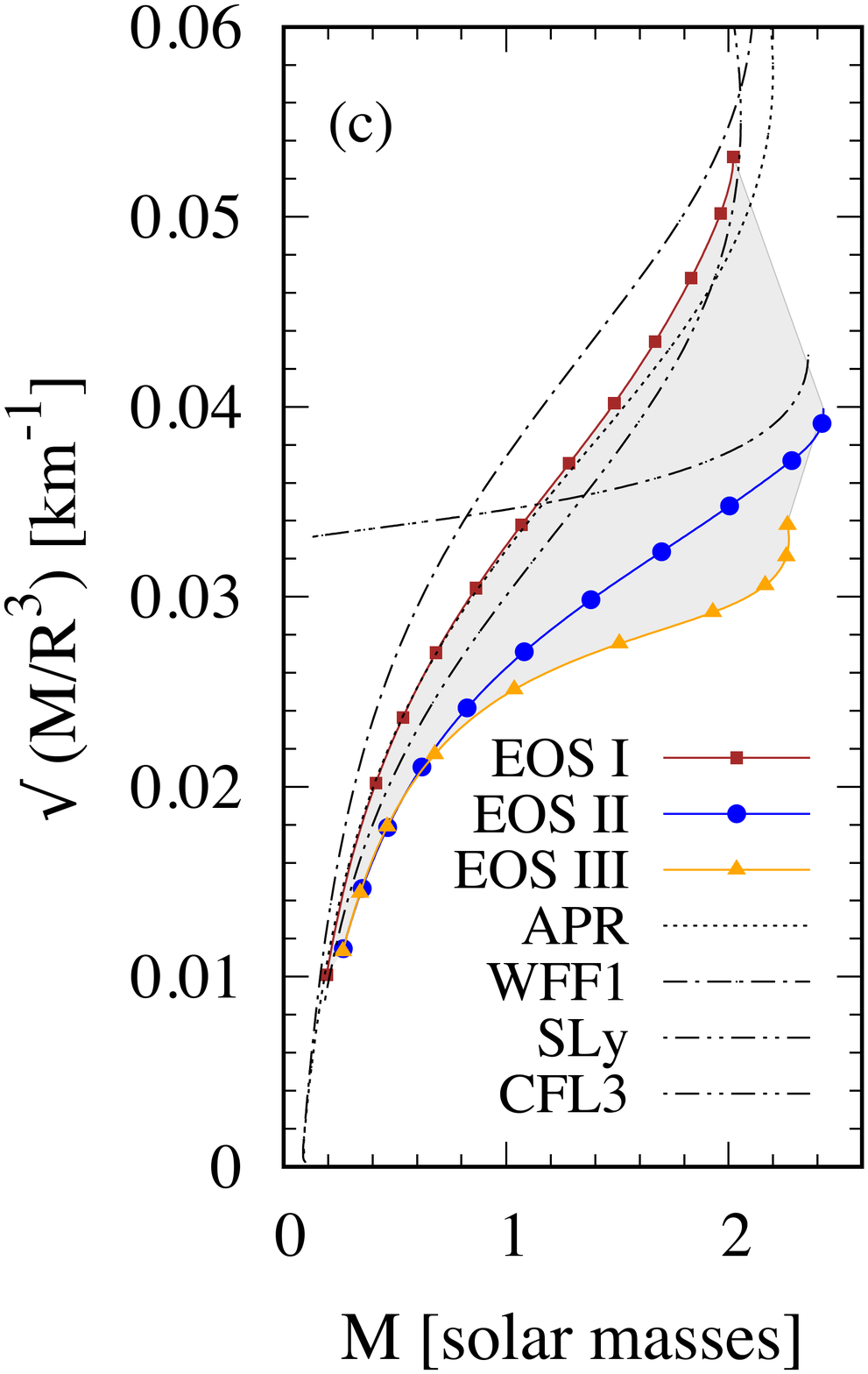}
\caption{For the three EOS shown in Fig. \ref{fig1} we show: (a) the mass-radius relationship, (b) the compactness $M/R$ as a function of the stellar mass $M$, and (c) the square root of the average stellar density $\sqrt{M/R^3}$ as a function of $M$.  The shaded area is consistent with low density nuclear matter and perturbative QCD. }
\label{fig2}
\end{figure*}

Varying the polytropic parameters and the transition density over ranges limited only by causality, a band of EOSs is obtained that can be further constrained by the requirement that the EOS is able to support a two solar mass star (to be in agreement with recent mass determinations using pulsar timing \cite{demorest2010,antoniadis2013}).  The resulting band is largely unaffected by the nature of the assumed phase transition or by the introduction of a third interpolating monotrope. Here we adopt three representative EOSs presented in tabulated form in Ref. \cite{2014ApJ...789..127K} which are maximally different from each other: EOS I gives the minimum radius, EOS II the maximum mass, and EOS III the maximum radius for compact stars (see Figs. \ref{fig1} and \ref{fig2}). For comparison we also include in Fig. \ref{fig2} results obtained with representative EOS models for hadronic matter (WFF \cite{WFF}, APR \cite{1998PhRvC..58.1804A} and Sly \cite{SLY}) and strange quark matter in a color flavor locked state (model CFL3 of Ref. \cite{2017PhRvC..95b5808F}).

A similar strategy for the EOS has been presented in \cite{2013ApJ...773...11H}, using  the  the same low-density EOS as here and taking into account the 2 $M_{\odot}$ condition.
However,  Ref. \cite{2013ApJ...773...11H} doesn't require the result to approach the perturbative QCD EOS at large densities and therefore they obtain a somewhat wider band for the allowed EOSs.

In addition to the very stringent limits on the maximum mass of NSs imposed by pulsar timing measurements \cite{demorest2010,antoniadis2013}, there has been in recent years a considerable progress in the determination of stellar radii (see e.g. \cite{Ozel2016} and references therein).  For example, high-quality observation of the thermal emission from Low Mass X-ray Binaries (LMXB)  in quiescence and during X-ray bursts allowed the determination of the radii of 14 NSs. The results for different objects fall in the range $\sim 7-15$ km and, when combined,  indicate NS radii in the $9.8-11$ km range \cite{Ozel2016}. Gravitational waves from binary mergers can also provide valuable constrains on NS radii. Recently, LIGO/Virgo detected the event GW170817, a coalescence of a NS binary system \cite{abbott2017b}. Restricting the component spins to the range inferred in binary NSs,  the observed data constrained the component masses to be in the range $1.17 - 1.60 M_{\odot}$.  Requiring that the equation of state supports NSs with masses larger than $1.97 M_{\odot}$, the analysis allows to constrain the radii to the range $10.5- 13.3$ km at the 90\% credible level \cite{LIGO2018}. 
Not all of the models presented in Fig. \ref{fig2} are in agreement with current radii observations. However, we must keep in mind that radii determinations are not yet as robust as mass measurements using pulsar timing. Because of this reason, in this work we keep only the 2 $M_{\odot}$ condition for constraining the equation of state.

\section{The formalism for studying non-radial oscillations}
\label{formalism}

The equilibrium configuration of a non-rotating compact star composed of a perfect fluid is determined by the Tolman-Oppenheimer-Volkoff equations
\begin{eqnarray}
\label{tov1}
\frac{dp}{dr} &=& - \frac{\epsilon m}{r^2}\bigg(1 + \frac{p}{\epsilon}\bigg)
	\bigg(1 + \frac{4\pi p r^3}{m}\bigg)\left(1 - \frac{2m}{r}\right)^{-1},  \\
\label{tov2}
\frac{d\nu}{dr} &=& - \frac{2}{\epsilon} \frac{dp}{dr} \bigg(1 + \frac{p}{\epsilon}\bigg)^{-1}, \\
\label{tov3}
\frac{dm}{dr}& =& 4 \pi r^2 \epsilon,
\end{eqnarray}
where  $m$ is the gravitational mass inside the radius $r$  and $\nu(r)$ is a metric potential. The pressure $p$ and the mass-energy density $\epsilon$ are related by the equations of state given before. The boundary conditions are $m(0) =0$, $p(R) =0$ and  $\nu(R)= \ln ( 1- {2M}/{R} )$,  where $R$ is the radius of the star and $M$ its mass.

The polar non-radial perturbations of a non-rotating star can be described through a set of equations presented in \cite{1983ApJS...53...73L, 1985ApJ...292...12D}.  The perturbed metric tensor  reads
\begin{eqnarray}
ds^2 & = & -e^{\nu}(1+r^{\ell}H_0Y^{\ell}_{m}e^{i\omega t})dt^2  - 2i\omega r^{\ell+1}H_1Y^{\ell}_me^{i\omega t}dtdr + e^{\lambda}(1 - r^{\ell}H_0Y^{\ell}_{m}e^{i \omega t})dr^2 \nonumber \\
&& + r^2(1 - r^{\ell}KY^{\ell}_{m}e^{i \omega t})(d\theta^2 + \sin^2\theta d\phi^2),
\end{eqnarray}
and the polar perturbations in the fluid are given  by the following Lagrangian displacements
\begin{eqnarray}
\xi^{r} &=& r^{\ell-1}e^{-\lambda/2}WY^{\ell}_{m}e^{i\omega t}, \\
\xi^{\theta} &=& -r^{\ell - 2}V\partial_{\theta}Y^{\ell}_{m}e^{i\omega t}, \\
\xi^{\phi} &=& -r^{\ell}(r \sin \theta)^{-2}V\partial_{\phi}Y^{\ell}_{m}e^{i\omega t} ,
\end{eqnarray}
where $Y^{\ell}_{m}(\theta,\phi)$ are the spherical harmonics, and we restrict to the $l = 2$ component, which dominates the emission of gravitational waves.

With this, non-radial oscillations are described by the following set of first order linear differential equations \cite{1985ApJ...292...12D}:
\begin{eqnarray}
H_1' &=&  -r^{-1} [ \ell+1+ 2Me^{\lambda}/r +4\pi   r^2 e^{\lambda}(p-\epsilon)] H_{1}  +  e^{\lambda}r^{-1}  \left[ H_0 + K - 16\pi(\epsilon+p)V \right] ,      \label{osc_eq_1}  \\
 K' &=&    r^{-1} H_0 + \frac{\ell(\ell+1)}{2r} H_1   - \left[ \frac{(\ell+1)}{r}  - \frac{\nu'}{2} \right] K    - 8\pi(\epsilon+p) e^{\lambda/2}r^{-1} W \:,  \label{osc_eq_2} \\
 W' &=&  - (\ell+1)r^{-1} W   + r e^{\lambda/2} [ e^{-\nu/2} \gamma^{-1}p^{-1} X  - \ell(\ell+1)r^{-2} V + \tfrac{1}{2}H_0 + K ] \:,  \label{osc_eq_3} \\
X' &=&  -  \frac{\ell}{r} X + \frac{(\epsilon+p)e^{\nu/2}}{2}  \Bigg[ \left( \frac{1}{r} + \frac{\nu'}{2} \right)H_{0}   + \left(r\omega^2e^{-\nu} + \frac{\ell(\ell+1)}{2r} \right) H_1    + \left(\tfrac{3}{2}\nu' - \frac{1}{r} \right) K   \nonumber  \\
&&    - \ell(\ell+1)r^{-2}\nu' V  -  \frac{2}{r}   \Biggl( 4\pi(\epsilon+p)e^{\lambda/2}  + \omega^2e^{\lambda/2-\nu}  - \frac{r^2}{2}  (e^{-\lambda/2}r^{-2}\nu')' \Biggr) W \Bigg]  \:,
\label{osc_eq_4}
\end{eqnarray}
where the prime denotes a derivative with respect to $r$ and  $\gamma$ is the adiabatic index. The function $X$ is given by
\begin{eqnarray}
X =  \omega^2(\epsilon+p)e^{-\nu/2}V - \frac{p'}{r}e^{(\nu-\lambda)/2}W  
  + \tfrac{1}{2}(\epsilon+p)e^{\nu/2}H_0 ,
\end{eqnarray}
and $H_{0}$ fulfills the algebraic relation 
\begin{eqnarray}
a_1  H_{0}= a_2  X -  a_3 H_{1}  + a_4 K,
\end{eqnarray}
with
\begin{eqnarray}
a_1 &=&  3M + \tfrac{1}{2}(l+2)(l-1)r + 4\pi r^{3}p  ,  \\
a_2 &=&  8\pi r^{3}e^{-\nu /2}   , \\
a_3 &=&  \tfrac{1}{2}l(l+1)(M+4\pi r^{3}p)-\omega^2 r^{3}e^{-(\lambda+\nu)}  ,  \\
a_4 &=&  \tfrac{1}{2}(l+2)(l-1)r - \omega^{2} r^{3}e^{-\nu}  -r^{-1}e^{\lambda}(M+4\pi r^{3}p)(3M - r + 4\pi r^{3}p)   .
\end{eqnarray}
Outside the star, the perturbation functions that describe the motion of the fluid vanish and the  system of differential equations reduces to the Zerilli equation:
\begin{equation}
\frac{d^{2}Z}{dr^{*2}}=[V_{Z}(r^{*})-\omega^{2}]Z,
\end{equation}
where $Z(r^{*})$ and $dZ(r^{*})/dr^{*}$  are related to the metric perturbations  $H_{0}(r)$ and $K(r)$ by transformations given in Refs. \cite{1983ApJS...53...73L,1985ApJ...292...12D}.  The ``tortoise'' coordinate is $r^{*} = r + 2 M \ln (r/ (2M) -1)$, and  the effective potential  $V_{Z}(r^{*})$ is given by
\begin{eqnarray}
V_{Z}(r^{*}) = \frac{(1-2M/r)}{r^{3}(nr + 3M)^{2}}[2n^{2}(n+1)r^{3} + 6n^{2}Mr^{2}  
 + 18nM^{2}r + 18M^{3}],
\end{eqnarray}
with $n= (l-1) (l+2) / 2$.

The system of Eqs. (\ref{osc_eq_1})$-$(\ref{osc_eq_4}) has four linearly independent solutions for given values of $l$ and $\omega$.  The physical solution needs to verify the appropriate boundary conditions.  
\textit{(a)} The perturbation functions must be finite everywhere, in particular at $r = 0$ where the non-radial oscillation equations are singular. To implement such condition  it is necessary to use a power series expansion of the solution near the singular point $r=0$ (the procedure is explained in detail in \cite{1983ApJS...53...73L}. \textit{(b)} The Lagrangian perturbation in the pressure has to be zero at the surface of the star $r = R$. This implies that the function $X$ must vanish at $r = R$.  Given specific values $l$ and $\omega$,   there is a unique solution that satisfies the above boundary conditions inside the star.

In general, outside the star the perturbed metric describes a combination of outgoing and ingoing gravitational waves.  The physical solution of the Zerilli equation is the one that describes purely outgoing gravitational radiation at $r=\infty$. 
Such boundary condition cannot be verified by any value of $\omega$ and the frequencies that fulfill this requirement represent the quasinormal modes of the stellar model. The numerical procedure to solve the above  equations is explained in \cite{1983ApJS...53...73L, 1985ApJ...292...12D}.

A variety of modes can be studied through the  above equations. The so called purely gravitational modes ($w$-modes) do not induce fluid motion and are highly damped. The so called fluid modes ($f$, $g$ and $p$) are usually classified according to the source of the restoring force which prevails in bringing the perturbed element of fluid back to the equilibrium position, e.g. buoyancy in the case of $g$-modes or a gradient of pressure for  $p$-modes. 
The frequencies of $g$-modes are lower than those of $p$-modes, and the two sets are separated by the frequency of the fundamental ($f$) mode. 
The \textit{f}-mode eigenfunctions have no nodes inside the star, and they grow towards the surface. Its frequency has been shown to be quite proportional to the square root of the mean density of the star and tends to be independent of the details of the stellar structure. 
In this work we focus only on the $f$ mode because it is expected to be the most excited in astrophysical events and therefore the main contribution to the GW emission of the star should be expected at its frequency.

\section{Results}
\label{results}

We integrated the stellar structure equations  \eqref{tov1}$-$\eqref{tov3} for EOS I, II, III and obtained the results shown in Fig. \ref{fig2}. As stated previously, of all models that are consistent with low density nuclear matter and perturbative QCD, EOS I gives the minimum stellar radius, EOS II the maximum mass, and EOS III the maximum stellar radius. Thus, the shaded area in the $M-R$ plane contains essentially all the stellar models that are consistent with the required constraints (see Ref. \cite{2014ApJ...789..127K} for a detailed discussion).

\begin{figure*}[tb]
\centering
\includegraphics[angle=-90,scale=0.30]{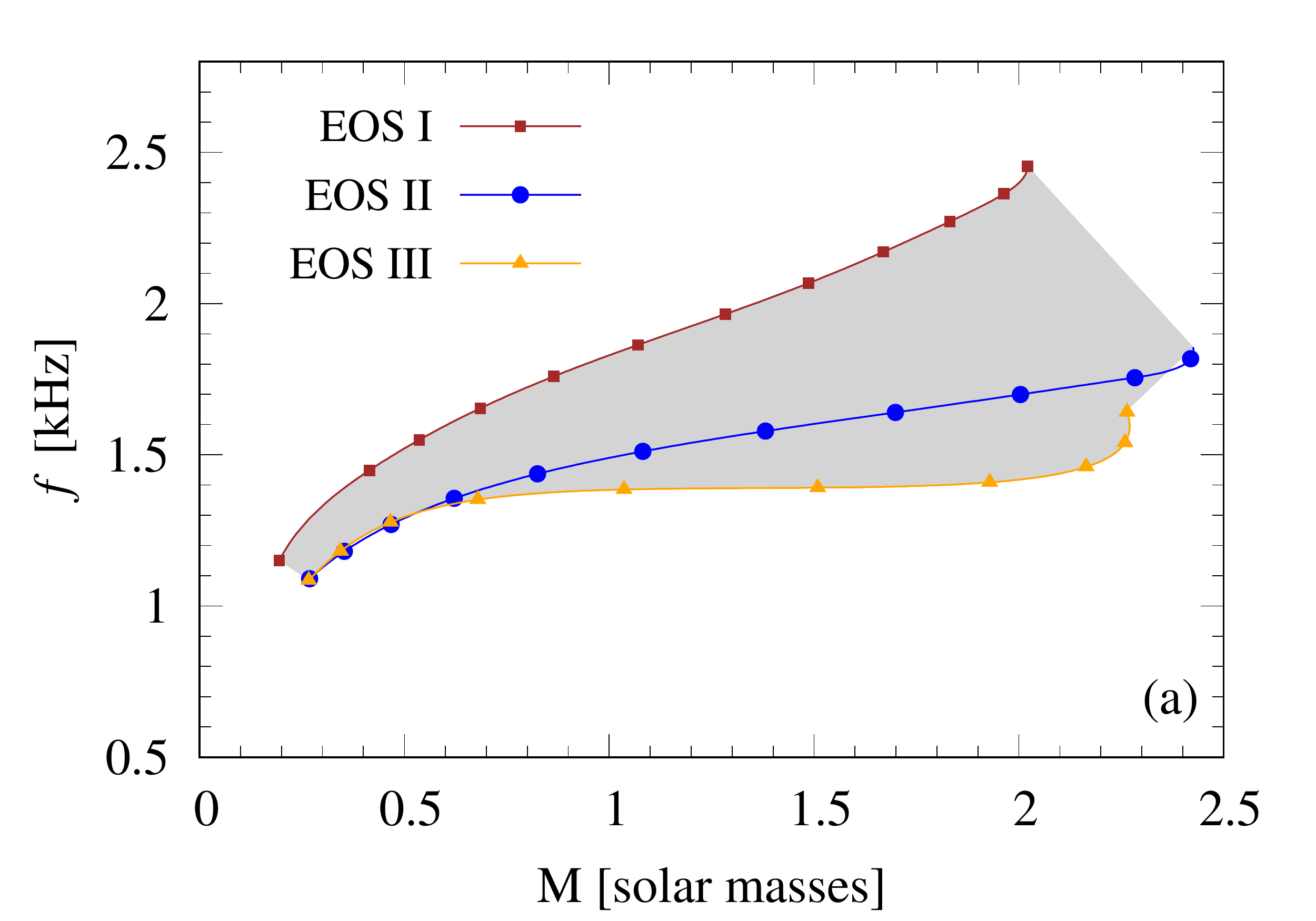}
\includegraphics[angle=-90,scale=0.30]{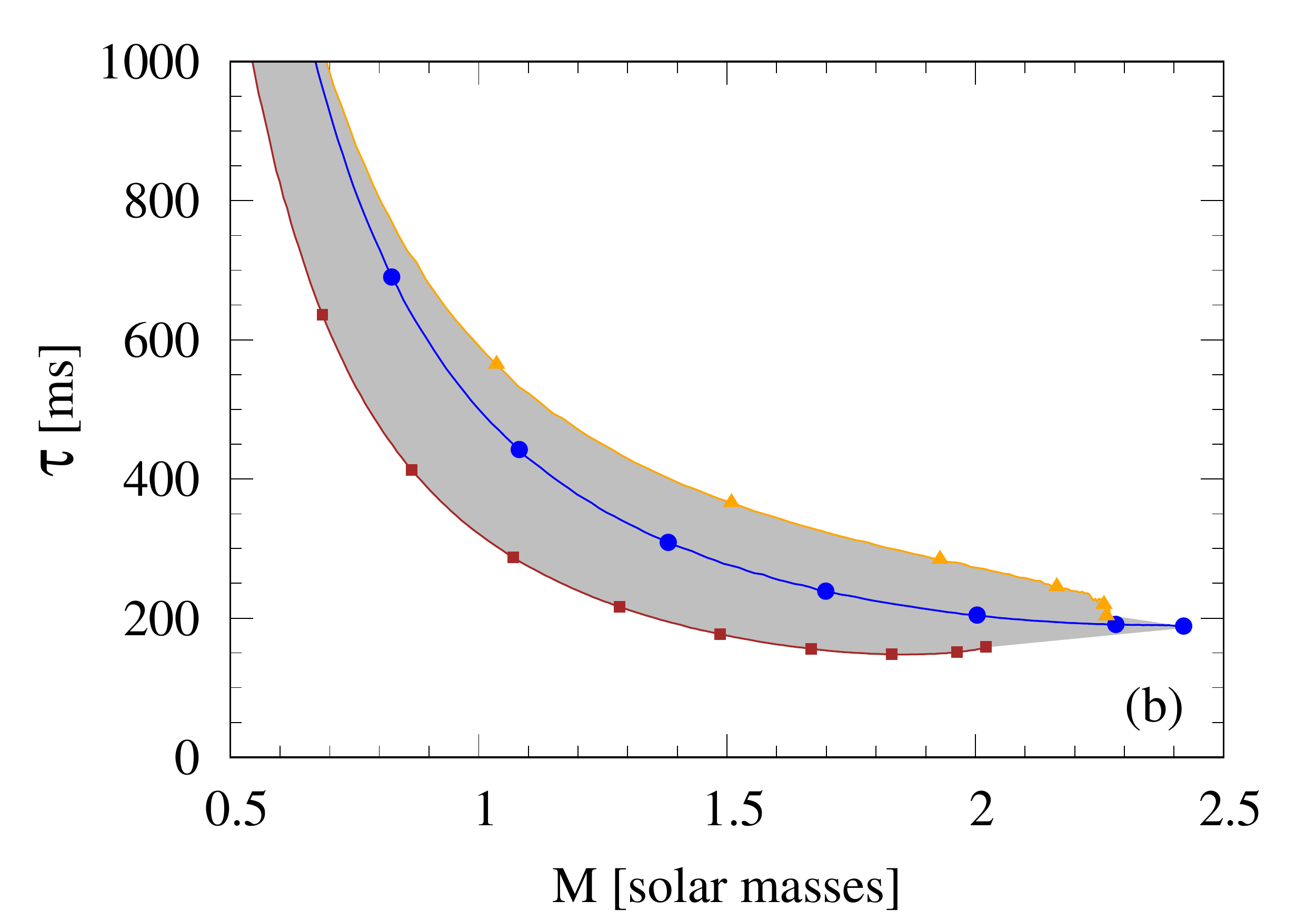}
\caption{Frequency and damping time of the fundamental mode as a function of the stellar mass, for EOS I, II and III. The shaded area is consistent with low density nuclear matter and perturbative QCD.} 
\label{fig3}
\end{figure*}

\begin{figure*}[tb]
\centering
\includegraphics[angle=-90,scale=0.30]{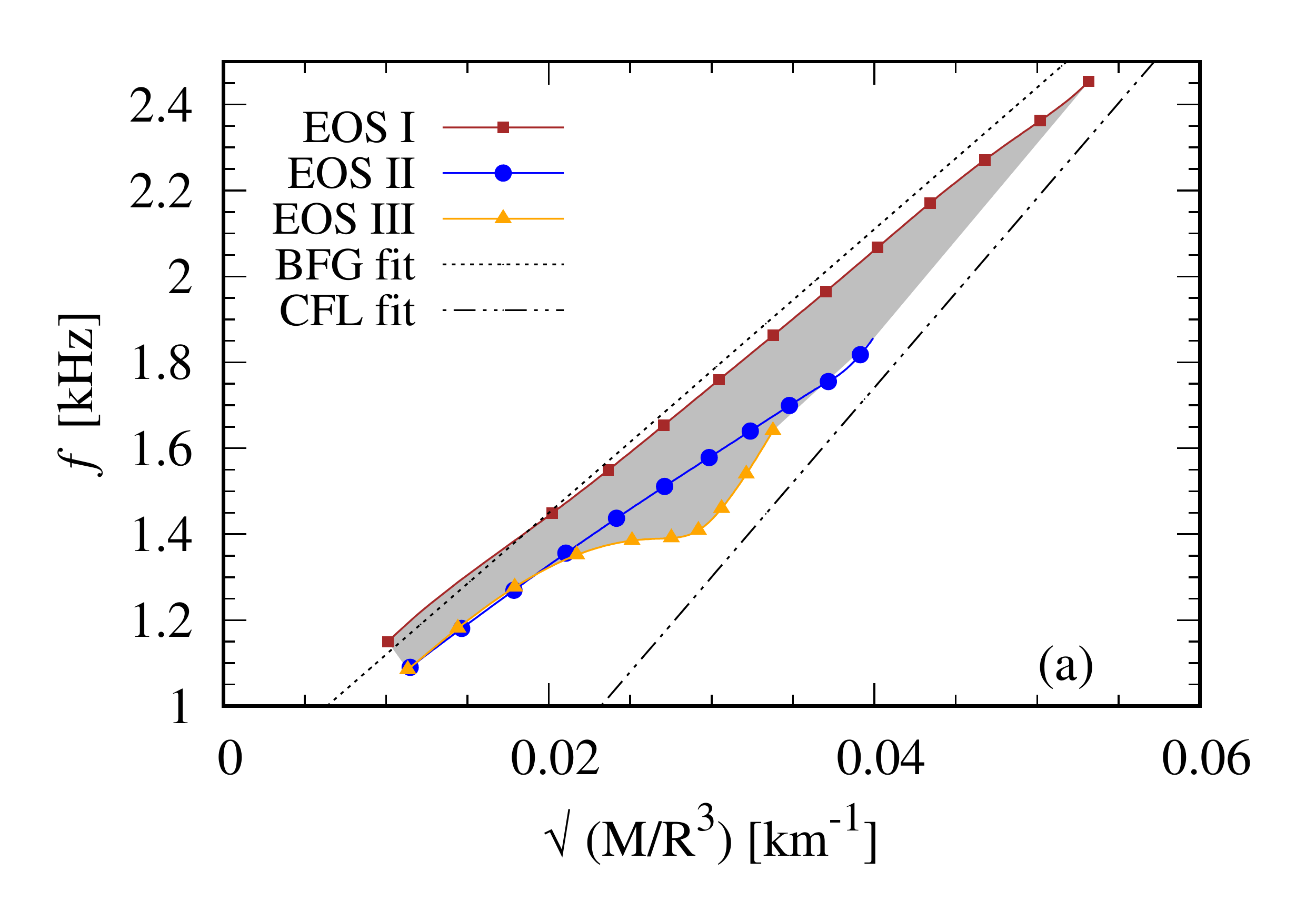}
\includegraphics[angle=-90,scale=0.30]{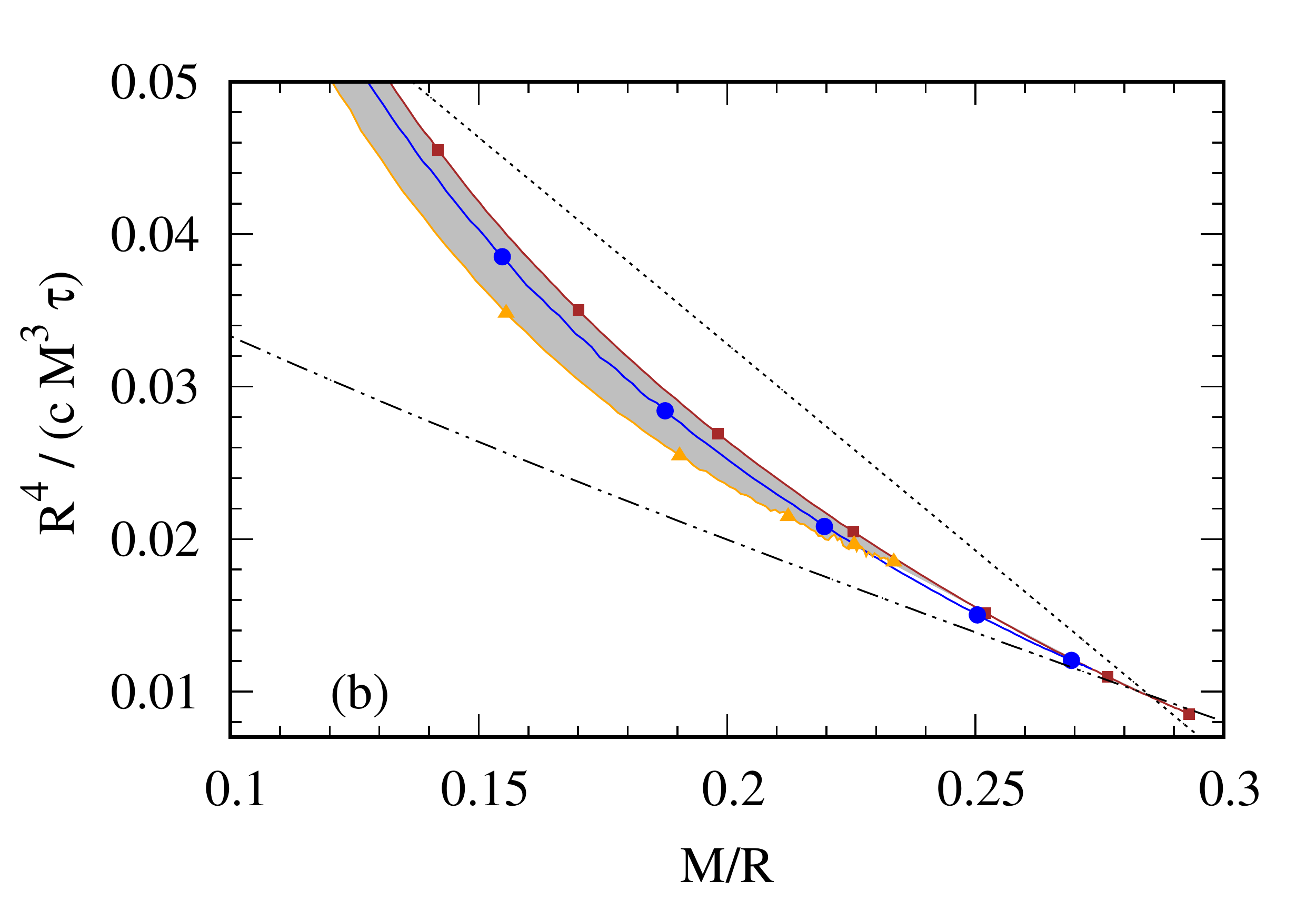}
\caption{The frequency of the fundamental mode as a function of the square root of the average density and the damping time as a function of the compactness $M/R$. For the frequency and damping time, we also plot the fittings of Benhar et al. \cite{Benhar:2004xg} (BFG fit), and the fit for color-flavor locked (CFL) strange quark matter given in \cite{2017PhRvC..95b5808F}   (CFL fit).}
\label{fig4}
\end{figure*}

The oscillation equations given in the previous section are integrated for quadrupole modes $(l=2)$ using the method explained in Ref. \cite{ 2017PhRvC..95b5808F}. This procedure allows obtaining $\omega$ for each value of the central density of the star, or equivalently for each value of the stellar mass.  The real part of $\omega$  is the pulsation frequency ($f = \mathrm{Re}(\omega)/2 \pi$) and the imaginary part is the inverse of the damping time of the mode due to gravitational-wave emission ($\tau= 1/\mathrm{Im}(\omega)$).

On the left panel of Fig. (\ref{fig3}), we show the  frequency $f$ of the fundamental mode for stars with different masses. For NS masses above $\sim 1 M_{\odot}$, which is the range of astrophysical interest, the oscillation frequency falls within a narrow window in the range $\sim 1.4-2.4$ kHz. 
In particular, for a canonical $1.4 M_{\odot}$ neutron star the maximum frequency is 2 kHz and for $2 M_{\odot}$ it is 2.4 kHz. 
The corresponding values for the damping time are showed on the right panel of Fig. (\ref{fig3}). 
All these results are in good agreement with a large amount of results obtained in the last years using a variety of phenomenological models. Still, we must emphasize that the main contribution of the present results is that they provide an EOS-independent band where future $f$-mode measurements should be expected.   

An interesting and potentially useful aspect of the $f$-mode is that it can be described by universal fitting formulae that are argued to be quite EOS-independent \cite{anderssonkokkotas1998,Benhar:2004xg,Chirenti:2015dda,Lau:2009bu}. In fact, it has been shown that the oscillation frequency $f$  of the fundamental mode has a reasonably linear dependence on the square root of the average density and the damping time $\tau$  can be fitted with simple formulae involving  the stellar mass and radius. Using several phenomenological hadronic EOSs   it has been shown in Ref. \cite{anderssonkokkotas1998} that  general ``empirical relations'' can be obtained for both $f$ and $\tau$. Later, new empirical relations were obtained employing additional hadronic and quark EOSs \cite{Benhar:2004xg,Flores:2013yqa}, different fitting formulae \cite{Chirenti:2015dda,Lau:2009bu},  and models for absolutely stable quark matter \cite{2017PhRvC..95b5808F}. These results suggest that the mass and radius of a compact object could be inferred if  $f$ and $\tau$ were detected by the new generation of gravitational wave detectors. However, there is in general some spread of the phenomenological EOSs around the fitting curves.
Thus, it would be desirable to establish reliable boundaries, as EOS-independent as possible,   to the expected properties of the $f$-mode. To this end, we have explored the same empirical relations analyzed in \cite{anderssonkokkotas1998,Benhar:2004xg} using the EOSs presented in Sec. \ref{EOS}.
On the left panel of Fig. (\ref{fig4}) we show $f$ as a function of the square root of the average density. For low average densities, which according to Fig. \ref{fig2} correspond to low-mass stars, our results are close to the BFG fit for hadronic stars \cite{Benhar:2004xg}, but for large average densities (high mass stars)  the curves fall near the fit corresponding to CFL strange quark stars \cite{2017PhRvC..95b5808F}. We observe a similar behavior for $\tau$ in the right panel of Fig. (\ref{fig4}).
The band obtained for the EOS (see Sec. \ref{EOS}), translates into well defined  windows for $f$ and $\tau$ as functions of combinations of the stellar mass and radius. Since these windows are narrow and rather model independent, the simultaneous observation  $f$ and $\tau$ can  be used to set robust constrains on the mass and radius of compact objects. 

For applications it is convenient to have analytic fittings of the allowed bands for $f$ and $\tau$.
The upper ($f_{\uparrow}$) and lower ($f_{\downarrow}$) limits of the shaded region on the left panel of Fig. \ref{fig4} can be approximated by:
\begin{equation}
f_{\uparrow} [\textrm{kHz} ]   = a_0 + a_1 x    \quad  \textrm{for}  \quad  0.01 <  x  <   0.053 
\end{equation}
where $x = \sqrt{M/R^3}$ in km$^{-1}$,  $a_0 = 0.83$,  $a_1= 30.6$,  and
\begin{equation}
f_{\downarrow}[\textrm{kHz} ]  =
  \begin{cases}
      b_0 + b_1 x + b_2 x^2        &  \textrm{for} \quad  0.011 <  x  <   0.0244  \\
      c_0 + c_1 x + c_2 x^2   &   \textrm{for} \quad  0.0244 <  x  <   0.0333  \\
      d_0 + d_1 x    &   \textrm{for} \quad  0.0333 <  x  <   0.04 
  \end{cases}
\end{equation}
with $b_0 = 0.512$,     $b_1= 62.5 $ ,      $b_2 = -1106$, $c_0 = 4.49$,     $c_1= -241$ ,     $c_2 = 4648$, $d_0 = 0.49   $, $d_1= 33.9$.
The allowed region for $R^4/(c M^3 \tau)$ is very narrow, specially for high compactness $M/R$. Thus, we provide a single fitting curve for the whole region:
\begin{equation}
R^4/(c M^3 \tau) = q_0 \left(\frac{M}{R}\right)^{q_1} + q_2 \left(\frac{M}{R}\right)^{q_3}
\end{equation}
with $q_0 = -0.05$, $q_1 = 0.83$, $q_2 = 0.008$ and $q_3 = -0.97$.  


Finally, let us consider the detectability of the fundamental pulsation mode, which depends crucially on  the level of excitation in an astrophysical situation. Here we will estimate what amount of energy should be channelled through the $f-$mode in order for it to be detectable by current and planned GW observatories. Following \cite{Echevarria1989,Apostolatos2001} we assume that the signal $h(t)$ emitted by the NS can be modeled as a damped sinusoid $h(t) = {\cal A} e^{- t/ \tau} \sin[2 \pi f t]$, where ${\cal A}$ is the initial amplitude of the signal. 
The energy flux $F$ carried by a weak gravitational wave $h(t)$ is given by $F= c^3| {\dot h} |^2 / (16 \pi G)$ where $c$ is the speed of light and $G$ is the Newton's constant.  Thus, when GWs emitted from an oscillating NS arrive at a detector on Earth, their initial amplitude will be: 
\begin{equation}
{\cal A} \sim 2.4 \times 10^{-20} \left( \frac{E_{\rm gw}}{10^{-6} M_{\odot} c^2}  \right)^{1/2} \left( \frac{10 {\rm kpc} }{D}  \right) \left( \frac{1 {\rm kHz}}{f}  \right) \left( \frac{1 {\rm
ms} }{\tau} \right)^{1/2}
\label{detection1}
\end{equation}
where $E_{gw}$ is the energy released through the $f-$mode, and $D$ is the distance between the source and the detector.

On the other hand, the signal-to-noise ratio at a detector is found to be \cite{Echevarria1989,Apostolatos2001}
\begin{equation}
\left( \frac{S}{N} \right)^2 = \frac{4 Q^2}{1+4 Q^2} \; \frac{{\cal A}^2 \tau}{2 S_n} ,
\label{detection2}
\end{equation}
where $Q \equiv \pi f \tau$ is the quality factor of the oscillation and $S_n$ is the noise power spectral density of the detector (assumed to be constant over the bandwidth of the signal).

From Eqs. \eqref{detection1} and \eqref{detection2} it is found:
\begin{equation}
 \left( \frac{E_{\rm gw}}{M_{\odot} c^2}  \right)  = 3.47 \times 10^{36} \left( \frac{S}{N} \right)^2   \frac{1+4 Q^2}{4 Q^2}  \left( \frac{D}{10 {\rm kpc} }  \right)^2 \left( \frac{f}{1 {\rm kHz}}  \right)^2 \left( \frac{S_n}{1 {\rm Hz^{-1}} } \right) .
\label{detection3}
\end{equation}

Let us focus on two different detectors: one of them with $S_n^{1/2} \sim 2 \times 10^{-23} \, \mathrm{Hz}^{-1/2}$ which is representative of the Advanced LIGO/Virgo at $\sim$kHz \cite{abbott2017b}, and  the other one with $S_n^{1/2} \sim 10^{-24} \, \mathrm{Hz}^{-1/2}$ which is illustrative of the planned third-generation ground-based Einstein Observatory at the same frequencies \cite{Einstein}. 
Let us consider a ``typical'' stellar model (see Fig. \ref{fig3}) for which the fundamental mode has $f=1.5$ kHz and $\tau=300$ ms.  
In this case we find that for a NS at the Virgo cluster ($D \sim 15$ Mpc) the energy channeled in the $f-$mode must be $E_{gw} >4\times 10^{-4} \, M_{\odot}c^2$ to have a signal-to-noise ratio $S/N>5$ in the Einstein telescope, and $E_{gw}>0.2 \, M_{\odot}c^2$ to have $S/N>5$ in LIGO/Virgo. 
For a NS in our Galaxy ($D \sim 10$ kpc) we must have $E_{gw}>2 \times 10^{-10} \, M_{\odot}c^2$ to lead to S/N>5 in the Einstein telescope, and
$E_{gw}>0.8 \times 10^{-7} \, M_{\odot}c^2$ to have S/N>5 in LIGO/Virgo.
The total energy estimated to be radiated as GWs in a core collapse supernova is $\sim 10^{-5}-10^{-6} \, M_{\odot}c^2$. Thus,  we can expect the detection of GWs from pulsating NSs within our own Galaxy, but not from far beyond it.

\section{Summary and conclusions}
\label{conclusions}

In the present paper, we studied the fundamental mode of non-radial oscillations of non-rotating compact stars using  a set of  EOSs  obtained from interpolation between the regimes of low-density nuclear matter and high-density perturbative QCD.  These two limiting cases are the outcome of reliable calculations within the  fundamental theory of  strong interactions and, as a consequence, they represent robust limits that should be fulfilled by the EOS at intermediate densities.
These two limits are connected by interpolating monotropes that are all subluminal.
Additionally, the resulting EOS must be able to support a star with  $2 M_{\odot}$, in agreement with the confirmed existence of  two solar mass pulsars. 
Since there is some stress between the softness imposed by the perturbative EOS at high densities and the stiffness required by the $2 M_{\odot}$ condition, the  allowed EOSs at intermediate densities fall within a narrow stripe (see Fig. \ref{fig1}) which in turn determines a  well defined region in the mass-radius diagram of compact stars (see Fig. \ref{fig2}). 
In the present work we used three different EOSs presented in Ref. \cite{2014ApJ...789..127K}:  EOS I and EOS III  were selected because they delimit the left and right boundaries of the allowed region in the mass-radius plane and EOS II was chosen because it gives the largest possible mass, $\sim 2.5 M_{\odot}$  (see Fig. \ref{fig2}).

Using EOSs I, II and III, we have solved the oscillation equations for the fundamental quadrupole mode and determined  the  frequency $f$ and the damping time $\tau$  for NSs with different masses. 
For stellar masses above $\sim 1 M_{\odot}$, which is the range of astrophysical interest, the oscillation frequency falls within a narrow window in the range $\sim 1.4-2.4$ kHz. For the same mass interval, the damping time varies between $200-600 \, \mathrm{ms}$ (see Fig. \ref{fig3}). 
We have also explored whether our results for the $f$-mode properties can be described through universal fitting formulae already known from the literature \cite{anderssonkokkotas1998,Benhar:2004xg}. Specifically, we calculated $f$ as a function of the square root of the average density and $R^4/(c M^3 \tau)$ as a function of the NS compactness $M/R$. 
In the case of the frequency, we find that  for low mass NSs the curves fall close to the hadronic fit presented in Ref. \cite{Benhar:2004xg} but for high mass NSs they approach to the fit for CFL strange quark matter given in Ref. \cite{2017PhRvC..95b5808F}) (see Fig. \ref{fig4}). A similar behavior is observed for the damping time. We provide simple analytic formulae of the allowed bands for $f$ and $\tau$.

As emphasized previously, our calculations are in good agreement with many previous works that used  a variety of phenomenological EOSs. 
Notwithstanding, the results presented here are important because they constrain the possible values for $f$ and $\tau$ within quite model independent windows that are the consequence of state-of-the-art reliable microphysics.
In principle, detections of the $f$-mode of \textit{non-rotating} compact stars should fall  within the windows presented in Figs. \ref{fig3} and \ref{fig4}. 
Measurements falling outside these windows, could indicate that some fundamental aspect not included in the present EOSs might play a significant role. 
For example,  there is  a family of exotic compact stars that cannot be described by the mere  interpolation between state-of-the-art low and high density EOSs (e.g. CFL strange quark stars entirely composed by color superconducting quark matter \cite{2017PhRvC..95b5808F}, and in general any model for strange quark matter). 
Additionally, in the case of hybrid stars composed by hadronic and quark matter separated by a sharp interface, a significant role may be played by the speed of the phase conversion of oscillating fluid elements in the neighborhood of the quark-hadron interface \cite{Pereira:2017rmp}. 
This emphasizes once more the necessity of gravitational wave astrophysical measurements to set bounds on the properties of ultra dense matter.

\acknowledgments
C.V.F is grateful to Comiss\~ao de Aperfei\c coamento de Pessoal do N\'ivel Superior (CAPES) of the Brazilian government. G.L. acknowledges the Brazilian agencies
Conselho Nacional de Desenvolvimento Cient\'{\i}fico e Tecnol\'ogico (CNPq) and Funda\c c\~ao de Amparo \`a Pesquisa do Estado de S\~ao Paulo (FAPESP) for financial support.


\begin{thebibliography}{10}

\bibitem{2014ApJ...789..127K}
A.~{Kurkela}, E.~S. {Fraga}, J.~{Schaffner-Bielich}, and A.~{Vuorinen},
  ``{Constraining Neutron Star Matter with Quantum Chromodynamics},'' {\em
  Astrophys. J.}, vol.~789, p.~127, July 2014.

\bibitem{abbott2017b}
B.~P. Abbott, R.~Abbott, T.~D. Abbott, {\em et~al.}, ``G{W}170817: Observation
  of gravitational waves from a binary neutron star inspiral,'' {\em Phys. Rev.
  Lett.}, vol.~119, p.~161101, Oct 2017.

\bibitem{flanagan2008}
E.~E. Flanagan and T.~Hinderer, ``Constraining neutron-star tidal love numbers
  with gravitational-wave detectors,'' {\em Phys. Rev. D}, vol.~77, p.~021502,
  Jan 2008.

\bibitem{read2009}
J.~S. Read, C.~Markakis, M.~Shibata, K.~b.~o. Ury\ifmmode~\bar{u}\else
  \={u}\fi{}, J.~D.~E. Creighton, and J.~L. Friedman, ``Measuring the neutron
  star equation of state with gravitational wave observations,'' {\em Phys.
  Rev. D}, vol.~79, p.~124033, Jun 2009.

\bibitem{abbott2017c}
B.~P. Abbott, R.~Abbott, T.~D. Abbott, {\em et~al.}, ``Gravitational waves and
  gamma-rays from a binary neutron star merger: {GW}170817 and {GRB}
  170817{A},'' {\em The Astrophysical Journal Letters}, vol.~848, no.~2,
  p.~L13, 2017.

\bibitem{pian2017}
E.~Pian, P.~D'Avanzo, S.~Benetti, {\em et~al.}, ``{Spectroscopic identification
  of r-process nucleosynthesis in a double neutron star merger},'' {\em
  Nature}, vol.~551, pp.~67--70, 2017.

\bibitem{shibata1994}
M.~Shibata, ``{Effects of tidal resonances in coalescing compact binary
  systems},'' {\em Prog. Theor. Phys.}, vol.~91, pp.~871--884, 1994.

\bibitem{parisi2017}
A.~Parisi and R.~Sturani, ``{Gravitational waves from neutron star excitations
  in binary inspirals},'' {\em arXiv:1705.04751}, 2017.

\bibitem{bauswein2016}
A.~Bauswein, N.~Stergioulas, and H.-T. Janka, ``{Exploring properties of
  high-density matter through remnants of neutron-star mergers},'' {\em Eur.
  Phys. J.}, vol.~A52, no.~3, p.~56, 2016.

\bibitem{andersson2011}
N.~Andersson, V.~Ferrari, D.~I. Jones, {\em et~al.}, ``{Gravitational waves
  from neutron stars: Promises and challenges},'' {\em Gen. Rel. Grav.},
  vol.~43, pp.~409--436, 2011.

\bibitem{faber2012}
J.~A. Faber and F.~A. Rasio, ``{Binary Neutron Star Mergers},'' {\em Living
  Rev. Rel.}, vol.~15, p.~8, 2012.

\bibitem{rezzolla2013}
L.~Rezzolla and O.~Zanotti, {\em Relativistic hydrodynamics}.
\newblock Oxford University Press, 2013.

\bibitem{Camelio:2017nka}
G.~Camelio, A.~Lovato, L.~Gualtieri, O.~Benhar, J.~A. Pons, and V.~Ferrari,
  ``{Evolution of a proto-neutron star with a nuclear many-body equation of
  state: Neutrino luminosity and gravitational wave frequencies},'' {\em Phys.
  Rev.}, vol.~D96, no.~4, p.~043015, 2017.

\bibitem{Warszawski:2012zq}
L.~Warszawski and A.~Melatos, ``{Gravitational-wave bursts and stochastic
  background from superfluid vortex avalanches during pulsar glitches},'' {\em
  Mon. Not. Roy. Astron. Soc.}, vol.~423, p.~2058, 2012.

\bibitem{lugones2002}
G.~Lugones, C.~R. Ghezzi, E.~M. de~Gouveia Dal~Pino, and J.~E. Horvath,
  ``{Asymmetric combustion in neutron stars and a potential mechanism for
  gamma-ray bursts},'' {\em Astrophys. J.}, vol.~581, pp.~L101--L104, 2002.

\bibitem{abdikamalov2008}
E.~B. Abdikamalov, H.~Dimmelmeier, L.~Rezzolla, and J.~C. Miller,
  ``{Relativistic simulations of the phase-transition-induced collapse of
  neutron stars},'' {\em Mon. Not. Roy. Astron. Soc.}, vol.~394, pp.~52--76,
  2009.

\bibitem{1967ApJ...149..591T}
K.~{Thorne} and A.~{Campolattaro}, ``Non-radial pulsation of
  general-relativistic stellar models. i. analytic analysis for l >= 2,'' {\em
  ApJ}, vol.~149, p.~591, 1967.

\bibitem{1968ApJ...152..673T}
K.~S. {Thorne} and A.~{Campolattaro}, ``Erratum: Non-radial pulsation of
  general-relativistivc stellar models. i. analytic analysis for l >= 2,'' {\em
  ApJ}, vol.~152, p.~673, 1968.

\bibitem{1983ApJS...53...73L}
L.~{Lindblom} and S.~L. {Detweiler}, ``{The quadrupole oscillations of neutron
  stars},'' {\em Astrophys. J. Suppl. Ser.}, vol.~53, pp.~73--92, Sept. 1983.

\bibitem{2013ApJ...773...11H}
K.~{Hebeler}, J.~M. {Lattimer}, C.~J. {Pethick}, and A.~{Schwenk}, ``{Equation
  of State and Neutron Star Properties Constrained by Nuclear Physics and
  Observation},'' {\em Astrophys. J.}, vol.~773, p.~11, Aug. 2013.

\bibitem{2014ApJ...781L..25F}
E.~S. {Fraga}, A.~{Kurkela}, and A.~{Vuorinen}, ``{Interacting Quark Matter
  Equation of State for Compact Stars},'' {\em Astrophysics. J. Lett.},
  vol.~781, p.~L25, Feb. 2014.

\bibitem{1998PhRvC..58.1804A}
A.~{Akmal}, V.~R. {Pandharipande}, and D.~G. {Ravenhall}, ``{Equation of state
  of nucleon matter and neutron star structure},'' {\em Phys. Rev. C}, vol.~58,
  pp.~1804--1828, Sept. 1998.

\bibitem{2013PhRvC..87a4322C}
L.~{Coraggio}, J.~W. {Holt}, N.~{Itaco}, R.~{Machleidt}, and F.~{Sammarruca},
  ``{Reduced regulator dependence of neutron-matter predictions with
  perturbative chiral interactions},'' {\em Phys. Rev. C}, vol.~87, p.~014322,
  Jan. 2013.

\bibitem{2012PhRvC..85c2801G}
S.~{Gandolfi}, J.~{Carlson}, and S.~{Reddy}, ``{Maximum mass and radius of
  neutron stars, and the nuclear symmetry energy},'' {\em Phys. Rev. C},
  vol.~85, p.~032801, Mar. 2012.

\bibitem{2013PhRvC..87a4338H}
J.~W. {Holt}, N.~{Kaiser}, and W.~{Weise}, ``{Chiral Fermi liquid approach to
  neutron matter},'' {\em Phys. Rev. C}, vol.~87, p.~014338, Jan. 2013.

\bibitem{2010PhRvC..82a4314H}
K.~{Hebeler} and A.~{Schwenk}, ``{Chiral three-nucleon forces and neutron
  matter},'' {\em Phys. Rev. C}, vol.~82, p.~014314, July 2010.

\bibitem{2012PhRvC..86e4317S}
F.~{Sammarruca}, B.~{Chen}, L.~{Coraggio}, N.~{Itaco}, and R.~{Machleidt},
  ``{Dirac-Brueckner-Hartree-Fock versus chiral effective field theory},'' {\em
  Phys. Rev. C}, vol.~86, p.~054317, Nov. 2012.

\bibitem{2013PhRvL.110c2504T}
I.~{Tews}, T.~{Kr{\"u}ger}, K.~{Hebeler}, and A.~{Schwenk}, ``{Neutron Matter
  at Next-to-Next-to-Next-to-Leading Order in Chiral Effective Field Theory},''
  {\em Phys. Rev. Lett.}, vol.~110, p.~032504, Jan. 2013.

\bibitem{demorest2010}
P.~Demorest, T.~Pennucci, S.~Ransom, M.~Roberts, and J.~Hessels, ``{Shapiro
  Delay Measurement of A Two Solar Mass Neutron Star},'' {\em Nature},
  vol.~467, pp.~1081--1083, 2010.

\bibitem{antoniadis2013}
J.~Antoniadis {\em et~al.}, ``{A Massive Pulsar in a Compact Relativistic
  Binary},'' {\em Science}, vol.~340, p.~6131, 2013.

\bibitem{1977PhRvD..16.1169F}
B.~A. {Freedman} and L.~D. {McLerran}, ``{Fermions and gauge vector mesons at
  finite temperature and density. III. The ground-state energy of a
  relativistic quark gas},'' {\em Phys. Rev. D}, vol.~16, pp.~1169--1185, Aug.
  1977.

\bibitem{1978PhRvD..17.2092B}
V.~{Baluni}, ``{Non-Abelian gauge theories of Fermi systems:
  Quantum-chromodynamic theory of highly condensed matter},'' {\em Phys. Rev.
  D}, vol.~17, pp.~2092--2121, Apr. 1978.

\bibitem{2005PhRvD..71j5014F}
E.~S. {Fraga} and P.~{Romatschke}, ``{Role of quark mass in cold and dense
  perturbative QCD},'' {\em Phys. Rev. D}, vol.~71, p.~105014, May 2005.

\bibitem{2010PhRvD..81j5021K}
A.~{Kurkela}, P.~{Romatschke}, and A.~{Vuorinen}, ``{Cold quark matter},'' {\em
  Phys. Rev. D}, vol.~81, p.~105021, May 2010.


\bibitem{WFF} R. B. Wiringa, V. Fiks, and A. Fabrocini, {\em Phys. Rev. C} 38, 1010 (1988).


\bibitem{SLY}  F. Douchin and P. Haensel,  {\em Astron. Astrophys} 380, 151 (2001).


\bibitem{2017PhRvC..95b5808F}
C.~V. {Flores} and G.~{Lugones}, ``{Constraining color flavor locked strange
  stars in the gravitational wave era},'' {\em Phys. Rev. C}, vol.~95,
  p.~025808, Feb. 2017.


\bibitem{Ozel2016} F. Ozel and P. Freire, "Masses, Radii, and the Equation of State of Neutron Stars " {\em Annu. Rev. Astron. Astrophys.} 54, 401 (2016).


\bibitem{LIGO2018}  The LIGO Scientific Collaboration and The Virgo Collaboration, "GW170817: Measurements of neutron star radii and equation of state", arXiv:1805.11581, (2018). 


\bibitem{1985ApJ...292...12D}
S.~{Detweiler} and L.~{Lindblom}, ``{On the nonradial pulsations of general
  relativistic stellar models},'' {\em Astrophys. J.}, vol.~292, pp.~12--15,
  May 1985.

\bibitem{Benhar:2004xg}
O.~Benhar, V.~Ferrari, and L.~Gualtieri, ``{Gravitational wave asteroseismology
  revisited},'' {\em Phys. Rev.}, vol.~D70, p.~124015, 2004.



\bibitem{anderssonkokkotas1998}
N.~Andersson and K.~D. Kokkotas, ``{Towards gravitational wave
  asteroseismology},'' {\em Mon. Not. Roy. Astron. Soc.}, vol.~299,
  pp.~1059--1068, 1998.

\bibitem{Chirenti:2015dda}
C.~Chirenti, G.~H. de~Souza, and W.~Kastaun, ``{Fundamental oscillation modes
  of neutron stars: validity of universal relations},'' {\em Phys. Rev.},
  vol.~D91, no.~4, p.~044034, 2015.

\bibitem{Lau:2009bu}
H.~K. Lau, P.~T. Leung, and L.~M. Lin, ``{Inferring physical parameters of
  compact stars from their f-mode gravitational wave signals},'' {\em
  Astrophys. J.}, vol.~714, pp.~1234--1238, 2010.

\bibitem{Flores:2013yqa}
C.~V. Flores and G.~Lugones, ``{Discriminating hadronic and quark stars through
  gravitational waves of fluid pulsation modes},'' {\em Class. Quant. Grav.},
  vol.~31, p.~155002, 2014.

\bibitem{Pereira:2017rmp}
J.~P. Pereira, C.~V. Flores, and G.~Lugones, ``{Phase transition effects on the
  dynamical stability of hybrid neutron stars},'' {\em Astrophys. J.} 860, 12 (2018).
  
\bibitem{Echevarria1989} F. Echevarria,  ``Gravitational-wave measurements of the mass and angular momentum of a black hole'', {\em Phys. Rev. D} 40, 3194 (1989). 

\bibitem{Apostolatos2001} K. D. Kokkotas, T. A. Apostolatos and N. Andersson, ``The inverse problem for pulsating neutron stars: a `fingerprint analysis' for the supranuclear equation of state'', {\em Mon. Not. R. Astron. Soc.} 320, 307 (2001).

\bibitem{Einstein} B. P. Abbott et al., ``Exploring the sensitivity of next generation gravitational wave detectors'' {\em Class. Quantum Grav.} 34,  044001 (2017).

\end{thebibliography}

\end{document}